\documentstyle[amssymb,aps,prb,preprint]{revtex}

\begin{document}
\draft
\author{John F. Dobson and Hung M. Le}
\title{Collisionless hydrodynamics for 1D motion of inhomogeneous degenerate
electron gases:\ equivalence of two recent descriptions}
\address{School of Science and Research Centre for Modelling and\\
Computation\\
Griffith University, Nathan, Queensland 4111, Australia}
\date{16 Jan 2002}
\maketitle

\begin{abstract}
Recently I. Tokatly and O. Pankratov (''TP'', Phys. Rev. B 60, 15550 (1999)
) used velocity moments of a classical kinetic equation to derive a
hydrodynamic description of electron motion in a degenerate electron gas.
Independently, the present authors (Theochem 501-502, 327 (2000)) used
considerations arising from the Harmonic Potential Theorem (Phys. Rev. Lett.
73, 2244 (1994)) to generate a new form of high-frequency hydrodynamics for
inhomogeneous degenerate electron gases (HPT-N3 hydrodynamics). We show here
that TP hydrodynamics yields HPT-N3 hydrodynamics when linearized about a
Thomas-Fermi groundstate with one-dimensional spatial inhomnogeneity.
\end{abstract}




\section{Introduction}

The prediction of collective oscillations in degenerate electron gases is a
very old problem that has been addressed over the years at many different
levels of theory. In the limit of an infinite {\em homogeneous} electron gas
it has been possible to apply relatively sophisticated microscopic theory
beyond the Random Phase (RPA) approximation\cite
{RichardsonAshcroftDynFxc3DEG}$^{,}$\cite{BohmContiTosiFxcTransv}. The main
current issue in the degenerate case of the uniform gas is the effect of
electron-electron correlations on the relevant response functions, leading
to modified dispersion relations at finite wavenumber, possibly presaging
quantum phase transitions. Interestingly, however, the homogeneous plasmons
both in two and three dimensions are completely determined by sum rules in
the zero-wavenumber limit, so that elementary hydrodynamic arguments give
the correct answers in this limit.

The situation is not so well advanced for {\em inhomogeneous} degenerate
electron gases, where high-level dynamic microscopic theory is unwieldy,
though the RPA has been carried through some time ago for simple-metal
surface geometries\cite{FeibelmanRvw}. Calculations of the RPA type are now
possible at the numerical level for molecules and possibly larger systems
via a number of quantum-chemistry-inspired packages. Some form of Time
Dependent Density Functional Theory can include the correlation physics
approximately\cite{absdampxc}$^{,}$\cite{LiebschEffectXcOnDParam}. All of
these microscopic approaches are rather computationally intensive, however.

In order to deal with long-wavelength response in a computationally simple
fashion, Bloch\cite{BlochHydroStoppingPower} and others very long ago
introduced a hydrodynamic style of treatment for degenerate electron
systems. Various workers\cite{InfluenceElChDistOnSrfPlasDispBennett}$^{,}$%
\cite{ForstmannStenschkeInterfacePLasmons}$^{,}$\cite{bartonhydro}$^{,}$\cite
{EguiluzHardwallSlabHydro}$^{,}$\cite{DasSarmaQuinn79HydroJellSurf}$^{,}$
have used this idea in linearized form with the assumption that the pressure
perturbation is proportional to the density perturbation. In the early
treatments, inhomogeneity of the gas was envisaged only in terms of edges
bounding an otherwise uniform gas,\ and these edges were treated via
postulated boundary conditions. Such approaches, often with a ``hard-wall''
boundary condition\cite{EguiluzHardwallSlabHydro}, have given a largely
reasonable description of collective plasmon modes in confined systems such
as thin metal slabs and quantum wells, with however some shortcomings to be
discussed below. Because plasmons are high-frequency phenomena, the
connection between pressure and density perturbations in a gas of uniform
density $n$, 
\begin{equation}
\delta p=m\beta ^{2}\delta n  \label{DpFromDnGen}
\end{equation}
had to be used with a nonstandard ''high-frequency'' value of the
coefficient $\beta $ derivable from microscopic Lindhard response theory,
namely 
\begin{equation}
\beta ^{2}=\frac{3}{5}v_{F}^{2}(n)=\frac{3}{5}\hbar ^{2}(3\pi
^{2}n)^{2/3}/m^{2}  \label{HighFrequBeta}
\end{equation}
instead of the low-frequency value

\begin{equation}
\beta ^{2}=\frac{1}{3}v_{F}^{2}(n)  \label{Low-FrequBeta}
\end{equation}
derivable from the static pressure-density relation for a degenerate
electron gas.

An obvious defect of the hydrodynamic approach is that it can be guaranteed
only to describe the ''classical'' size effects such as standing plasmons in
which a finite number of nodes of the electron density perturbation fit
across the thin dimension of an electron gas layer. By contrast, ''quantum''
size effects involve details of wavefunctions and are not necessarily given
by hydrodynamic theory: these size effects are related directly to
transitions between discrete quantum states due to spatial confinement of
the electrons. These two categories of mode are not fully distinct, however%
\cite{WLSJFD94}: for e xample, the sloshing or Kohn mode of the electron gas
in a parabolic quantum well can be regarded either as due to transitions
between two Harmonic Oscillator states, or as an odd combination of surface
plasmons\cite{SrfCollModesNNJ}. Nevertheless, it is clear that in general
there can be many more modes in a microscopic treatment than in the
corresponding hydrodynamic one.

Apart from the lack of the complete spectrum of quantum size-effect modes,
the hydrodynamic approach, as usually applied, has other drawbacks for
confined electron gases. Firstly, it has mostly been used under the
assumption that the electron gas is spatially homogeneous. This assumption
then requires a separate treatment of edge effects, via somewhat arbitrary
boundary conditions at the edge of the uniform-gas region. For example, the
hard-wall boundary conditions and uniform spatial profile frequently used to
analyze plasmons on charge-neutral electron gases of finite width\cite
{EguiluzHardwallSlabHydro} give a reasonable description of the 2D plasmon
and sloshing modes (equivalent to even and odd combinations of surface
plasmons) but not the multipole surface plasmons, which require a selvage of
finite width. For an electron gas under parabolic confinement (non-neutral
gas) on the other hand, the same hard-wall condition does not seem to be so
appropriate\cite{DempseyHalperinHydro}$^{,}$\cite{TokatlyPankHydroII}, and
``free'' boundary conditions have been suggested along with some further
assumptions..

The hydrodynamic approach is particularly tricky for plasmon modes that are
intrinsically tied to inhomogeneity, such as the multipole surface plasmons
of a neutral jellium surface. These modes depend intimately on the detailed
decay of the electron density at the edge, and this cannot be treated by a
simple, single boundary condition on a uniform gas. There have been attempts
within hydrodynamics to model the edge by a series of steps, each with its
own distinct but internally homogeneous density\cite
{HydroSurfPlasmonsSchwartzSchaich}, in which case additional boundary
conditions must be used. Hydrodynamic models have also been investigated
with a smoother but somewhat arbitrary surface density profile\cite
{InfluenceElChDistOnSrfPlasDispBennett}$^{,}$\cite
{InfluenceElDensProfileOnSrfPlasDispEguil}$^{,}$\cite
{HydroSurfPlasEguiluzQuinn}. The problem then is that the profile is not
internally generated, and in some of these approaches the pressure term was
not properly consistent with the chosen inhomogeneous density profile.

In connection with inhomogeneous or bounded degenerate gases, we have taken
the point of view\cite{jfdprl94}$^{,}$\cite{PlasmonsTHEOCHEM} that
hydrodynamics should be applied to a continuously varying model of edges (or
of inhomogeneity in general). In addition, one should use a smooth solution
of the groundstate problem that is consistent with the hydrodynamic
approximation chosen. Then arbitrary boundary conditions can be kept to a
minimum or removed altogether. In addition, the pressure term should be
properly adapted to the inhomogeneity. For plasmon applications on a finite
slab of electron gas, we tried the obvious inhomogeneous generalization of
the high-frequency Bloch approach by using a high-frequency inhomogeneous
pressure coefficient (\ref{HighFrequBeta}) $\beta ^{2}(n_{0}(\vec{r}))=\frac{%
3}{5}v_{F}^{2}(n_{0}(\vec{r}))$, where $n_{0}(\vec{r})$ is the
selfconsistent Thomas-Fermi groundstate density profile. This approach was
found\cite{jfdprl94}, however, to violate the Harmonic Potential Theorem
(HPT). This theorem is an extension of the Generalized Kohn Theorem\cite
{GenKohnTheorem} and states among other things that, for systems confined by
a harmonic external potential (e.g. the electron gas in a parabolic quantum
well), there exists a plasmon mode in which the inhomogeneous groundstate
density (and groundstate many-body wavefunction) oscillates rigidly at the
bare harmonic oscillator frequency. This violation was subsequently\cite
{VignaleAccFrameSumRule} discussed in a more general fashion, from the
viewpoint of invariance under transformation to an accelerated reference
frame\cite{jfdcio93-article}. In Eqs. (21)-(24) of the original HPT\ paper%
\cite{jfdprl94} a modification of collisionless Bloch hydrodynamics was
suggested in order to satisfy the HPT. That version of hydrodynamics was
later found to violate Newton's third law (momentum conservation), but this
was rectified to produce a theory that we here term the HPT-N3 hydrodynamics%
\cite{PlasmonsTHEOCHEM}. HPT-N3 theory was applied successfully\cite
{PlasmonsTHEOCHEM} to describe both the Kohn and standing-plasmon modes of a
parabolic quantum well gas. We have also been able\cite{MultSrfPlasWoABC} to
use HPT-N3 hydrodynamics to see multipole surface plasmons at the surface of
a near-neutral electron gas, without using additional boundary conditions.

It is this modified HPT-N3 hydrodynamic theory that will be discussed
further here and compared with another hydrodynamic theory. Specifically,
Tokatly and Pankratov (``TP'') in two recent papers (``TP1''\cite
{TokatlyPankratovHighFreqHydro1D}, ``TP2''\cite{TokatlyPankHydroII}) derived
a hydrodynamic theory valid both for low and high frequencies, by taking
moments of an assumed semiclassical kinetic equation. They did not, however,
explicitly examine the case of spatially inhomogeneous groundstates (other
than via assumed boundary conditions on a homogeneous gas). We will show
here that this TP theory agrees with the HPT-N3 theory in the collisionless
limit, when taken to second order in gradients and linearized about the
appropriate Thomas-Fermi groundstate of an electron gas with general
one-dimensional inhomogeneity. Since the two derivations of the pressure
term are quite different, this agreement tends to support the validity of
both theories for the case of 1D inhomogeneity.

The present work is organized as follows. In Section II a brief summary is
given of the HPT-N3 theory. In section III the collisionless TP equations
through second order are quoted for the simplified case of 1D spatial
variations. In Section IV these 1D TP equations are linearized about the
inhomogeneous TF groundstate and the result is shown to be precisely the
HPT-N3 theory\cite{PlasmonsTHEOCHEM}. Section V is devoted to discussion and
conclusions.

\section{Summary of linearized HPT-N3 hydrodynamics}

The exact equation of motion (Euler, Newton II) for the fluid velocity $\vec{%
u}(\vec{r},t)$ is 
\begin{equation}
mn(\frac{\partial }{\partial t}+\vec{u}.\vec{\nabla})\vec{u}=-\vec{\nabla}.%
{\bf P}+n\vec{F}  \label{NonlinEuler}
\end{equation}
where ${\bf P}(\vec{r},t)$ is the stress-pressure tensor (not known exactly
in general),
 $n(\vec{r},t)$ is the number  density,
  and $\vec{F}\,(\vec{r},t)$ is the (external plus internal) force
per electron due to interactions, usually treated via some form of
mean-field theory. The pressure force per electron, i.e. the force due to
streaming of particles from one fluid element into another, is $\vec{F}%
^{press}=-n^{-1}\vec{\nabla}.{\bf P}$. The form (\ref{NonlinEuler})
guarantees that the net force on the system due to pressure is zero:
specifically, the total system pressure force is 
\[
F^{press,tot}=\int n(\vec{r},t)\vec{F}^{press}(\vec{r},t)d\vec{r}=\int (-%
\vec{\nabla}.{\bf P})d\vec{r}\,=0 
\]
Linearizing (\ref{NonlinEuler}) about a non-streaming ($\vec{u}_{0}=0$)
equilibrium state we obtain 
\begin{equation}
\;mn_{0}(\vec{r})\frac{\partial }{\partial t}\vec{u}_{1}(\vec{r},t)=-\vec{%
\nabla}\cdot {\bf P}_{1}(\vec{r},t)+n_{1}(\vec{r},t)\vec{F}_{0}(\vec{r}%
)+n_{0}(\vec{r})\vec{F}_{1}(\vec{r},t)  \label{LinEulerGenPress}
\end{equation}

The approach introduced previously\cite{jfdprl94} in order to accommodate
the HPT in a Bloch-style inhomogeneous hydrodynamics was to use the fluid
displacement $\vec{x}\,$defined by 
\begin{equation}
\vec{x}(r,t)\equiv \int_{t_{0}}^{t}\vec{u}(\vec{r},t^{\prime })dt^{\prime }
\label{VectorFluidDisplt}
\end{equation}
to write the density perturbation $n_{1}$ as the sum of a compressive part $%
n_{1A}$ and a displacive part $\ n_{1B}$:

\begin{eqnarray}
n_{1}(\vec{r},t) &=&-\vec{\nabla}\cdot \left( n_{0}\left( \vec{r}\right) 
\vec{x}(\vec{r},t)\right)  \label{ExactDnLinFromXi} \\
&=&n_{1A}(\vec{r},t)+n_{1B}(\vec{r},t)  \label{DnEqDnAPlusDnB} \\
n_{1A}(\vec{r},t) &=&-n_{0}(\vec{r})\vec{\nabla}\cdot \vec{x}(\vec{r},t)
\label{DefDnA} \\
n_{1B}(\vec{r},t) &=&-\vec{x}(\vec{r},t)\cdot \vec{\nabla}n_{0}(\vec{r})
\label{DefDnB}
\end{eqnarray}
Equation (\ref{ExactDnLinFromXi}) is exact to linear order and can be
obtained from a time integration of the linearized continuity equation
(conservation of electrons). Eqs. (\ref{DnEqDnAPlusDnB})-(\ref{DefDnB}) are
also exact. Note that $n_{1B}$ represents the density perturbation that
would result if the groundstate density distribution were locally displaced
rigidly without compression, while $n_{1A}\,$represents the density
perturbation due to compression. For Kohn-mode motion in which the
groundstate is translated rigidly, $n_{1B}\,=n_{1}$ while $n_{1A}\,=0$
because there is no compression. By contrast, for oscillations about a
spatially homogeneous equilibrium state for which $\vec{\nabla}n_{0}=\vec{0}$%
, we have $n_{1A}=n_{1}\,$while $n_{1B}=0.$

Thus in order to satisfy the known Lindhard limit for high-frequency
response of a uniform gas, it was proposed\cite{jfdprl94} that $n_{1A}\,$be
associated in general with the high-frequency pressure coefficient $\beta
^{2}(\infty )=\frac{3}{5}v_{F}^{2}$. Similarly to obtain the correct HPT\
behavior for inhomogeneous harmonically-confined systems it was further
proposed\cite{jfdprl94} that $n_{1B}$ be associated with the low-frequency
coefficient $\beta ^{2}(0)=\frac{1}{3}v_{F}^{2}\,$, because the static
groundstate pressure profile must be rigidly translated in Kohn-mode motion.
Then the pressure perturbation (assumed\cite{jfdprl94}$^{,}$\cite
{PlasmonsTHEOCHEM} to be a scalar rather than the more general tensor) is 
\begin{eqnarray}
p_{1}^{HPTN3}(\vec{r},t) &=&mv_{F}^{2}(n_{0}(\vec{r}))[\frac{3}{5}n_{1A}(%
\vec{r},t)+\frac{1}{3}n_{1B}(\vec{r},t)]  \label{DeltaPHPTN3} \\
P_{1ij} &=&\delta _{ij}p_{1}^{HPTN3}  \label{DiagonalP}
\end{eqnarray}
where 
\begin{equation}
v_{F}^{2}(n_{0}(\vec{r}))=\hbar ^{2}m^{-2}(3\pi ^{2}n_{0}(\vec{r}))^{2/3}
\label{LocVfSq}
\end{equation}
is the square of the local Fermi velocity corresponding to the groundstate
number density $n_{0}(\vec{r})$.$\,$Although motivated by the limiting
uniform and quadratically-confined cases, Eq. (\ref{DeltaPHPTN3}) was
proposed as a general result.

The linearized HPT-N3 theory is completed by inserting (\ref{DeltaPHPTN3})-(%
\ref{LocVfSq}) into the linearized Euler equation (\ref{LinEulerGenPress}).
For the situations of interest here where all space dependence is in the $z$
direction, this gives

\begin{equation}
0=mn_{0}\partial _{t}u_{1}-\partial _{z}(\frac{3}{5}mv_{F}^{2}n_{0}\partial
_{z}\xi +\frac{1}{3}mv_{F}^{2}\xi \partial _{z}n_{0})-n_{1}F_{0}-n_{0}F_{1}.
\label{HPTN31DLinEuler}
\end{equation}
Here $\xi \,$is the fluid displacement in the $z$ direction from (\ref
{VectorFluidDisplt}), 
\begin{equation}
\xi (z,t)=\int_{t_{0}}^{t}u_{1}(z,t^{\prime })dt^{\prime },
\label{1DFluidDispl}
\end{equation}
and $v_{F}$ is a position-dependent Fermi velocity determined by the
inhomogeneous groundstate density $n_{0}$: 
\begin{equation}
v_{F}^{2}=\hbar ^{2}m^{-2}(3\pi ^{2}n_{0}(z))^{2/3}.  \label{1DInhomVF}
\end{equation}

Equations (\ref{HPTN31DLinEuler}) - (\ref{1DInhomVF}) are the basic
equations of linear HPT-N3 hydrodynamics\cite{PlasmonsTHEOCHEM} for a
degenerate electron gas where both the groundstate and excitations exhibit
one-dimensional spatial inhomogeneity. They were derived so as to satisfy
the Harmonic Potential Theorem\cite{jfdprl94} (HPT), to obey conservation of
particle number and of momentum (hence Newton III, N3), and to give the
correct limit of high-frequency plasma dispersion in a uniform electron gas
as in microscopic degenerate response (Lindhard-RPA) theory. We will show
elsewhere\cite{JFDHLeUnpub02} that the HPT-N3 equations can alternatively be
derived from nonlinear time dependent density functional theory with a very
simple assumption regarding memory. The HPT-N3 equations represent an
improvement over the HPT\ hydrodynamics originally proposed\cite{jfdprl94},
in that the original form did not satisfy Newton III.

We will now outline the Tokatly and Pankratov (TP) hydrodynamic theory\cite
{TokatlyPankratovHighFreqHydro1D}$^{,}$\cite{TokatlyPankHydroII}, then
compare it with the HPT-N3 theory summarized above, for the case of
one-dimensional spatial variations and linear perturbations. Despite their
very different derivations, these two theories will be shown to agree for
this 1D case.

\section{The TP hydrodynamic moment equations for one-dimensional spatial
variation}

Tokatly and Pankratov\cite{TokatlyPankratovHighFreqHydro1D}$^{,}$\cite
{TokatlyPankHydroII} obtained hydrodynamic-style equations by taking
velocity moments of a classical kinetic equation, assuming that this
equation is adequate for degenerate Fermi particles, presumably with a fully
degenerate Fermi distribution in mind as the static solution.

They decoupled the set of hydrodynamic moment equations, using the smallness
of the parameter 
\begin{equation}
\gamma =\frac{<p/m>}{L\max \{\omega ,\nu _{c}\}}  \label{DefGamma}
\end{equation}
where $<p/m>$ is the average particle speed, $L\equiv q^{-1}$ measures the
spatial scale of the hydrodynamic perturbation, and $\nu _{c}$ is an
appropriate collision frequency. We are interested in the high-frequency,
collisionless case for which $\omega >>\nu _{c}$ and collision integrals can
be ignored as discussed by TP. To $O(\gamma ^{2})$, their first three 1D
moment equations (Eqs. (16), (17) and (18) of their first paper\cite
{TokatlyPankratovHighFreqHydro1D}) are then, in our notation, 
\begin{eqnarray}
m(\frac{\partial }{\partial t}+u\frac{\partial }{\partial z})n(z,t)+mn\frac{%
\partial }{\partial z}u &=&0  \label{TPContinuity} \\
\therefore \frac{\partial }{\partial t}n+\frac{\partial }{\partial z}(nu)
&=&0\,\,\,\,\,\,\,\,(continuity)  \nonumber
\end{eqnarray}
\begin{eqnarray}
mn(\frac{\partial }{\partial t}+u\frac{\partial }{\partial z})u+\frac{%
\partial }{\partial z}P-nF &=&0  \label{TPNewton2} \\
&&(Newton\;II)  \nonumber
\end{eqnarray}
\begin{eqnarray}
(\frac{\partial }{\partial t}+u\frac{\partial }{\partial z})P+3P\frac{%
\partial }{\partial z}u(z,t) &=&0\,\,\,\,  \label{TPPressureEvol} \\
&&(evolution\;of\;stress-pressure).  \nonumber
\end{eqnarray}
Here the number density $n$, fluid velocity $u,$ the stress-pressure tensor
component $P\equiv P_{zz},$ and the total force $F$ due to interactions, are
all functions of the single space coordinate $z$ and the time $t$. (The
three-dimensional generalization of (\ref{TPContinuity}) - (\ref
{TPPressureEvol}) is given by Equations (22) - (25) of TP2\cite
{TokatlyPankHydroII}, with ${\bf L}^{(3)}\,$and ${\bf I}^{(2)}$ set to
zero.) Our notation in (\ref{TPContinuity})-(\ref{TPPressureEvol}) is chosen
to match that of our recent work\cite{PlasmonsTHEOCHEM}. The connections
with the notation of TP1\cite{TokatlyPankratovHighFreqHydro1D} are as
follows, with our notation on the left: 
\begin{eqnarray*}
z &\rightarrow &x, \\
n &=&m^{-1}L^{(0)}, \\
u &=&v, \\
P &=&P_{xx}=L^{(2)}, \\
V &=&-e\phi , \\
F &=&+e\partial _{x}\phi .
\end{eqnarray*}

\section{Comparison of linearized HPT-N3 and TP theories}

In linearizing (\ref{TPContinuity}) - (\ref{TPPressureEvol}), Tokatly and
Pankratov immediately replaced the convective derivative $D_{t}\equiv
\partial _{t}+\vec{u}.\vec{\nabla}$ by the partial derivative $\partial _{t}$%
. (See Eqs. (23)-(26) of TP1\cite{TokatlyPankratovHighFreqHydro1D} and
Eqs.(46) - (49) ff. of TP2\cite{TokatlyPankHydroII}). In particular, this
implies that quantities such as $\partial P_{0}/\partial z$ and $\partial
n_{0}/\partial z\,$are set to zero, where $P_{0}(z)$ is the equilibrium
pressure-stress tensor component and $n_{0}(z)$ is the groundstate density
profile. Thus they specialized to systems with a spatially uniform
groundstate, without actually stating this condition. They thereby also
removed the possibility of treating edge inhomogeneities other than via
assumed boundary conditions. Because of the homogeneous assumption they were
furthermore unable to discuss the Harmonic Potential Theorem fully, even
though some reference was made to it in Section V of TP2\cite
{TokatlyPankHydroII}. Note that the essential feature of the HPT, which
makes it a more stringent test of many-particle theories compared with the
Generalized Kohn Theorem\cite{GenKohnTheorem} (GKT), is that the entire
groundstate many-body wavefunction, and hence all derived quantities such as 
$n_{0}(\vec{r})$ and $P_{0}(\vec{r})$, are rigidly translated in HPT motion,
giving $n(\vec{r},t)=n_{0}(r-X(t))$, $P(\vec{r},t)=P_{0}(r-X(t))$ etc.,
where the center-of-mass coordinate $X(t)$ satisfies the classical
simple-harmonic-oscillator equation of motion. To verify satisfaction of the
HPT, it is not sufficient to show that the centre of mass moves
appropriately as in TP2\cite{TokatlyPankHydroII}:\ the preservation of the
inhomogeneous groundstate spatial profiles $n_{0}(\vec{r})$, $P_{0}(\vec{r}%
)\,$in the moving situation is also an essential feature.

\subsection{Equilibrium state}

Fortunately it is easy to linearize (\ref{TPContinuity})-(\ref
{TPPressureEvol}) without discarding the space derivatives of groundstate
quantities. First note that, for a 3D electron gas in a 1D situation where
spatial variation of all quantities occurs only in the $z$ direction, the
kinetic equation for the equilibrium distribution function $f_{0}$ is 
\begin{equation}
\{m^{-1}p_{z}\partial _{z}+F_{0z}\frac{\partial }{\partial p_{z}}\}f_{0}(z,%
\vec{p})=0  \label{EquilmKinEqu}
\end{equation}
where $\vec{F}_{0}=F_{0z}\hat{z}$ is the total selfconsistent force due to
interactions. If $F_{0}(z)=-\partial _{z}V_{0}(z)$,$\,$then direct
substitution into (\ref{EquilmKinEqu}) verifies that the equilibrium
solutions are of the form 
\[
f_{0}(z,\vec{p})=A(\frac{1}{2m}\vec{p}^{2}+V_{0}(z)),\;\;\vec{p}^{2}\equiv
p_{x}^{2}+p_{y}^{2}+p_{z}^{2} 
\]
where $A(\varepsilon )\,$is an arbitrary function of one variable. For a
degenerate electron gas, the classical kinetic theory will optimally decribe
the quantal situation if $A\,$is chosen to be the Fermi distribution 
\begin{equation}
A(\varepsilon )=C\theta (\mu -\varepsilon ),
\end{equation}
where $C\,$and $\mu $ are independent of $z\,$and $\vec{p}$. The groundstate
density is then 
\[
n_{0}(z)=\int f_{0}(z,\vec{p})d^{3}p=C\int_{0}^{p_{F}(z)}4\pi p^{2}dp=C\frac{%
4\pi }{3}p_{F}^{3}(z) 
\]
where $C$ is a constant and 
\[
p_{F}(z)=mv_{F}(z)=\sqrt{2m(\mu -V_{0}(z)}. 
\]
Note that this assumption of a local Fermi distribution with a global
chemical potential amounts to a Thomas-Fermi type of theory, which is
therefore the natural groundstate theory to accompany a hydrodynamic
treatment of excitations in a degenerate system.

The $zz$ component of the stress-pressure tensor from this distribution (see
Eqs. (3) and (13) of TP1\cite{TokatlyPankratovHighFreqHydro1D})

is 
\begin{eqnarray}
P_{0zz} &\equiv &L_{0zz}=\frac{1}{m}\int p_{z}p_{z}f_{0}(\vec{r},\vec{p}%
)d^{3}p=\frac{1}{3}\frac{1}{m}\int \vec{p}^{2}f_{0}(z,\vec{p})d^{3}p 
\nonumber \\
&=&\frac{1}{3}\frac{1}{m}\int p^{2}C\theta (p_{F}(z)-p)4\pi p^{2}dp 
\nonumber \\
&=&\frac{1}{3m}\frac{4\pi }{5}p_{F}^{5}(z)=\frac{1}{5m}p_{F}^{2}(z)n_{0}(z)=%
\frac{m}{5}v_{F}^{2}(z)n_{0}(z)\propto n_{0}(z)^{5/3}  \label{P0zzEquilm}
\end{eqnarray}

Then 
\begin{equation}
\partial _{z}P_{0zz}\equiv \partial _{z}L_{0zz}^{(2)}=\partial _{z}n_{0}%
\frac{\partial P_{0zz}}{\partial n_{0}}=\partial _{z}n_{0}\frac{5}{3}\frac{%
P_{0zz}}{n_{0}}=\partial _{z}n_{0}\frac{1}{3m}p_{F}^{2}(z)=\partial _{z}n_{0}%
\frac{m}{3}v_{F}^{2}(z)  \label{DP0Dz}
\end{equation}

\subsection{Linearized moment equations of TP}

Linearizing our Eqs. (\ref{TPContinuity})-(\ref{TPPressureEvol}) about the
Thomas-Fermi groundstate just described, we write 
\begin{eqnarray*}
n(\vec{r},t) &=&n_{0}(z)+n_{1}(z,t) \\
u(\vec{r},t) &=&0+u_{1}(z,t) \\
P_{zz}(\vec{r},t) &=&P_{0zz}(z)+P_{1}(z,t) \\
F(\vec{r},t) &=&F_{0}(z)+F_{1}(z,t)
\end{eqnarray*}
and obtain

\begin{equation}
\;\partial _{t}n_{1}+n_{0}\frac{\partial u_{1}}{\partial z}=0
\label{LinZeroMom}
\end{equation}
\begin{equation}
\;mn_{0}\frac{\partial u_{1}}{\partial t}+\partial
_{z}P_{1}-n_{0}F_{1}-n_{1}F_{0}=0  \label{LinFirstMom}
\end{equation}
\begin{equation}
\partial _{t}P_{1}+u_{1}\partial _{z}P_{0zz}+3P_{0zz}\partial _{z}u_{1}=0
\label{LinSecondMom}
\end{equation}
where $n_{1}$, $u_{1}\,$and $P_{1}$ are functions of $z$ and $t$ whereas $%
n_{0}$, $F_{0}\,$and $P_{0zz}$ are functions of $z$ alone. Note that the $%
u_{1}\partial _{z}P_{0zz}$ term in (\ref{LinSecondMom}) is missing in (e.g.)
Eq. (49) of TP2 or Eq. (26) of TP1 because TP have assumed a uniform
unperturbed system so that $D_{t}=\partial _{t}.\,$It is this additional
term in our treatment which ensures satisfaction of the true HPT\ (not just
the GKT which is satisfied by the treatment of TP involving uniform gas with
zero-stress boundary condition).

Insertion of the equilibrium pressure tensor and its space derivative from (%
\ref{P0zzEquilm}) and (\ref{DP0Dz}) into (\ref{LinSecondMom}) gives 
\begin{eqnarray}
\partial _{t}P_{1}+u_{1}(\partial _{z}n_{0})\frac{m}{3}v_{F}^{2}(z)+\frac{3m%
}{5}v_{F}^{2}(z)n_{0}\partial _{z}u_{1} &=&0 \\
-\partial _{t}P_{1}=\frac{m}{3}v_{F}^{2}(z)u_{1}\partial _{z}n_{0}+ &&\frac{%
3m}{5}v_{F}^{2}(z)n_{0}\partial _{z}u_{1}.  \label{P1zzLinFromU1}
\end{eqnarray}
We integrate both sides with respect to time, up to time $t$ from an initial
equilibrium at time $t_{0}$ when all perturbations vanished. Then using (\ref
{1DFluidDispl}) we obtain 
\begin{equation}
P_{1}=-\frac{m}{3}v_{F}^{2}(z)\xi \partial _{z}n_{0}-\frac{3m}{5}%
v_{F}^{2}(z)n_{0}\partial _{z}\xi .  \label{P1FromXiTP}
\end{equation}
Inserting (\ref{P1FromXiTP}) into (\ref{LinFirstMom}) we obtain 
\begin{equation}
\;mn_{0}\frac{\partial u_{1}}{\partial t}-\partial _{z}(\frac{m}{3}%
v_{F}^{2}(z)\xi \partial _{z}n_{0}+\frac{3m}{5}v_{F}^{2}(z)n_{0}\partial
_{z}\xi )-n_{0}F_{1}-n_{1}F_{0}=0  \label{TP1DLinEuler}
\end{equation}

Eq. (\ref{TP1DLinEuler}) derived from the Tokatly-Pankratov approach\cite
{TokatlyPankratovHighFreqHydro1D} is identical to Eq. (\ref{HPTN31DLinEuler}%
) derived from the HPT-N3 theory\cite{PlasmonsTHEOCHEM}. This equivalence is
the principal result of the present paper. Eq. (\ref{TP1DLinEuler}) has
already been shown\cite{PlasmonsTHEOCHEM} to give a sensible treatment of
both the Kohn mode and the hydrodynamic standing plasmon modes of the
electron gas in a parabolic quantum well, while taking into account the
smooth decay of the electron density at the edges and using only ``natural''
boundary conditions that the fluid displacement $\xi $ and density
perturbation $n_{1}$ are nowhere divergent.

\section{Discussion}

We have compared two hydrodynamic theories describing plasmon excitations in
degenerate electron gases with one-dimensional spatial inhomogeneity. Their
derivations differ principally by the way the pressure term is obtained.

The first theory, our HPT-N3 hydrodynamics, obtains the pressure term by
identifying displacive and compressive components, $n_{1A}$ and $n_{1B}$, of
the density perturbation $n_{1}$: see Eqs. (\ref{ExactDnLinFromXi}) - (\ref
{DefDnB}). $n_{1A}$ is associated with the high-frequency pressure
coefficient $\beta ^{2}(\omega \rightarrow \infty )=\frac{3}{5}v_{F}^{2}$,
while $n_{1B}\,$is associated with the low-pressure coefficient $\beta
^{2}(\omega =0)=\frac{1}{3}v_{F}^{2}$. See Eq. (\ref{DeltaPHPTN3}). (We will
show elsewhere\cite{JFDHLeUnpub02} that these associations arise naturally
from a form of time-delayed scalar local density approximation for the
pressure.) By these means we were able to satisfy the Harmonic Potential
Theorem\cite{jfdprl94} (HPT) and the usual conservation laws, and to obtain
within the same formalism the correct plasmon dispersion of the uniform
electron gas. The HPT applies to harmonically confined systems and
constitutes a stringent test of an inhomogeneous many-particle theory,
requiring in the present case that there exist a Kohn mode in which the
inhomogeneous groundstate density profile $n_{0}(\vec{r})$ and pressure
profile $P_{0}(\vec{r})$ move rigidly. This linear HPTN3 theory has been
shown to give a sensible description\cite{PlasmonsTHEOCHEM} not only of the
Kohn mode and other (standing) modes of harmonically confined systems, but
also of multipole surface plasmons on a near-neutral jellium electron gas
layer\cite{JFDHLeUnpub02}.

The second theory in our comparison is that of Tokatly and Pankratov\cite
{TokatlyPankratovHighFreqHydro1D} (TP). This was derived for degenerate
systems by assuming the validity of a classical kinetic equation, then
truncating the momentum moment equations. This yields a prediction for the
pressure term. The truncation is justified, in the case of interest to us,
by the smallness of the parameter $\gamma =v_{F}/(L\omega )$ where $L\,\,$%
and $\omega $ are the spatial scale and frequency. In order to compare with
the HPT theory we needed to deal with inhomogeneous groundstates, and
therefore we linearized the TP\ theory about the appropriate inhomogeneous
Thomas-Fermi groundstate, a procedure not explicitly carried out by Tokatly
and Pankratov.

Despite their quite different derivations, these two linearized theories
give the same predictions for the case of one-dimensional spatial variation,
as evidenced by the identity of Eqs. (\ref{HPTN31DLinEuler}) and (\ref
{TP1DLinEuler}). This fact is the main result of the present paper.

The HPT-N3 derivation shows that the assumption of a classical kinetic
equation, made by TP even in degenerate cases, is not necessary. On the
other hand, the TP derivation is part of a systematic expansion whereas the
HPT-N3 derivation is not obviously part of any systematic scheme. Thus the
two approaches are somewhat complementary, and each tends to support the
validity of the other.\ 

In order to be truly useful, the inhomogeous formalisms discussed here
should ideally be applicable at edges where the groundstate and/or excited
electron density may vary rapidly in space. It is then likely that the TP
truncation parameter $\gamma \,$(Eq. (\ref{DefGamma})) is small only at
frequencies higher than the ones of interest. Nevertheless, in at least one
important case covered by the Harmonic Potential Theorem, the formalism
gives correct answers for very rapidly varying edge profiles. The use of
these hydrodynamic approaches in regimes of rapid spatial variation is
somewhat reminiscent of the commonplace and surprisingly successful use of
the Local Density Functional formalism\cite{KS} for groundstate properties
of highly inhomogeneous systems, despite the in-principle restriction to
slow spatial variation. In that case the success is at least partly
explained by the satisfaction of sum rules and constraints\cite
{JonesGunnarssonRMP}. The HPT constraint can be viewed in the same light for
time-dependent cases, and indeed has already been used for this purpose in
the context of the time-dependent Local Density Approximation for exchange
and correlation\cite{DBG97}. The connection of the present hydrodynamic
approximations with finite-memory versions of density functional theory will
be made more explicit elsewhere\cite{JFDHLeUnpub02}.

It should also be stressed that the existing HPT-N3 theory does not
necessarily agree with the more general three-dimensional version\cite
{TokatlyPankHydroII} of the TP theory. This is because the HPT-N3 theory to
date has assumed a scalar pressure, whereas in the high-frequency plasmon
case the pressure should certainly be a tensor, which indeed is what emerges
from the 3D TP\ theory. It will be interesting to see if a tensor ansatz for
the pressure, along the lines of the HPT-N3 argument, can re-derive the 3D
TP theory.

\bibliographystyle{prsty}
\bibliography{jfdall,hydro,dldf,statsrf,srfdynre,gendft}

\end{document}